\documentclass[sigconf, nonacm]{acmart}

\AtBeginDocument{%
  \providecommand\BibTeX{{%
    \normalfont B\kern-0.5em{\scshape i\kern-0.25em b}\kern-0.8em\TeX}}}

\copyrightyear{2023}
\acmYear{2023}
\acmConference[UbiComp/ISWC '23 GenAI4PC]{Generative AI for Pervasive Computing Symposium at the 2023 ACM International Joint Conference on Pervasive and Ubiquitous Computing \& the 2023 ACM International Symposium on Wearable Computing}{October 8--12, 2023}{Cancun, Quintana Roo, Mexico}

\usepackage{soul}

\begin{document}
\title[HAR using LLMs through Pose Estimation]{Towards Enhanced Human Activity Recognition through Natural Language Generation and Pose Estimation}
%\ninept

\author{Nikhil Kashyap}
\email{nkash@uw.edu}
%\orcid{}
\affiliation{%
  % \department{}
  \institution{University of Washington}
  \city{Seattle}
  \state{Washington}
  \country{United States}
}

\author{Manas Satish Bedmutha}
\email{mbedmutha@ucsd.edu}
\orcid{0000-0003-3427-2226}
\affiliation{%
  % \department{Computer Science and Engineering}
  \institution{University of California San Diego}
  \city{La Jolla}
  \state{California}
  \country{United States}
}

\author{Prerit Chaudhary, Brian Wood, Wanda Pratt, Janice Sabin, Andrea Hartzler}
\email{{prerit16, bwood2, wpratt, sabinja, andreah}@uw.edu}
%\orcid{}
\affiliation{%
  % \department{}
  \institution{University of Washington}
  \city{Seattle}
  \state{Washington}
  \country{United States}
}

% \author{Poorva Satish Bedmutha}
% \email{pbedmutha@ucsd.edu}
% \orcid{0000-0003-4640-0782}
% \affiliation{%
%   \department{Electrical and Computer Engineering}
%   \institution{University of California San Diego}
%   \city{La Jolla}
%   \state{California}
%   \country{United States}
% }

\author{Nadir Weibel}
\email{weibel@ucsd.edu}
\orcid{0000-0002-3457-4227}
\affiliation{%
% \department{Computer Science and Engineering}
  \institution{University of California San Diego}
  \city{La Jolla}
  \state{California}
  \country{United States}
}

\renewcommand{\shortauthors}{Kashyap et. al.}

\begin{abstract}
Vision-based human activity recognition (HAR) has made substantial progress in recognizing predefined gestures but lacks adaptability for emerging activities. This paper introduces a paradigm shift by harnessing generative modeling and large language models (LLMs) to enhance vision-based HAR. We propose utilizing LLMs to generate descriptive textual representations of activities using pose keypoints as an intermediate representation. Incorporating pose keypoints adds contextual depth to the recognition process, allowing for sequences of vectors resembling text chunks, compatible with LLMs. This innovative fusion of computer vision and natural language processing holds significant potential for revolutionizing activity recognition. A proof of concept study on a Kinetics700 dataset subset validates the approach's efficacy, highlighting improved accuracy and interpretability. Future implications encompass enhanced accuracy, novel research avenues, model generalization, and ethical considerations for transparency. This framework has real-world applications, including personalized gym workout feedback and nuanced sports training insights. By connecting visual cues to interpretable textual descriptions, the proposed framework advances HAR accuracy and applicability, shaping the landscape of pervasive computing and activity recognition research. As this approach evolves, it promises a more insightful understanding of human activities across diverse contexts, marking a significant step towards a better world.
\end{abstract}



\maketitle

\section{Introduction}
\label{sec:intro}

\begin{figure*}[t]
    \centering
    \includegraphics[width=\textwidth]{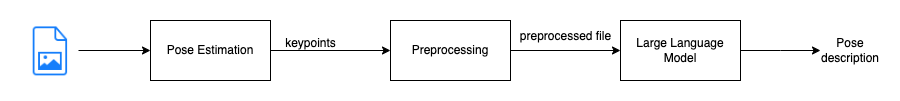}
   \Description[Overall Framework]{Proposed framework where each frame from a video feed can be used to first estimate the pose keypoints of the subjects, then preprocessed to use them with large language models}
   \caption{Proposed framework: Each frame from a video feed can be used to first estimate the pose keypoints of the subjects, then preprocessed to use them with large language models}
       \label{fig:framework} 
\end{figure*}

In the realm of pervasive computing, Human Activity Recognition (HAR) is an established  approach to attempt deciphering human actions and behaviors in various environments. However, despite numerous successful use-cases~\cite{dang2020sensor}, human activity recognition is still an evolving field and countless types of actions that we undergo on a daily basis still can't be recognized. As datasets corresponding to different activities keep growing~\cite{zhang2017review}, most current approaches develop models or pipelines catered to recognizing specific activities in those datasets. %Even models trained to generalize over a certain input modality are restricted to specific modalities as inputs. 
With recent advances in representational learning and generative modeling, we can leverage highly powerful large neural networks to adapt to changing needs of data, in terms of both the input applied and the outputs generated. In this work we explore the use of large language models as a solution to extend human activity recognition beyond vision-based approaches.

% Despite its potential, the current state of HAR techniques often falls short in accuracy, prompting the need for innovative solutions that can bridge the existing gaps. This paper delves into the domain of HAR, exploring its association with pervasive computing and delving into the intricacies of vision-based HAR methods. In this context, we introduce a novel framework that reimagines HAR as a Natural Language Generation (NLG) problem, proposing a paradigm shift in how human activities are interpreted and recognized.

\subsection{Challenges in Vision-Based HAR}
Vision-based HAR methods have garnered significant attention due to their potential for capturing intricate nuances of human actions, making them particularly valuable in applications where precision is paramount. For instance, these methods excel at discerning subtle hand movements in sign language interpretation, accurately tracking facial expressions in emotion recognition, and precisely capturing the body mechanics in sports actions like tennis serves or golf swings. However, these methods encounter several challenges that hinder their accuracy and practicality, ultimately limiting their effectiveness in real-world scenarios.

Traditional vision-based approaches often rely on complex models that necessitate substantial amounts of training data and meticulous parameter tuning. This dependency on large datasets introduces potential biases and restricts the applicability of the model to specific conditions. Additionally, the intricate architecture of these models can lead to computationally intensive operations, rendering them less feasible for deployment on resource-constrained devices or in real-time applications. Furthermore, the need for manual parameter adjustment makes the models less adaptable to new scenarios or activities without laborious reconfiguration.

In this work, we focus on addressing the issue of model generalization in vision-based HAR. Specifically, we tackle the challenge of adapting recognition models to real-world scenarios that exhibit variations in lighting conditions, camera viewpoints, and user demographics. This is a crucial aspect of our endeavor to enhance the accuracy and practicality of HAR methods, making them more robust and versatile in the face of dynamic and diverse environments.

\subsection{A Novel Approach - GroupFormer}
Recently, Li et. al.~\cite{li2021groupformer} proposed the use of language models to solve other vision based tasks through the GroupFormer framework.  It presents a unique perspective by leveraging both visual and human pose estimation features to enhance activity recognition. The model employs a pre-trained inflated 3D convolutional deep neural network (I3D-CNN) to extract visual features of individuals. These visual features are combined with pose estimation features from AlphaPose~\cite{fang2022alphapose}, resulting in a comprehensive set of player features. These player features serve as inputs to a stack of clustered spatial-temporal transformers, refining the representation of human activities.

GroupFormer, while promising, carries certain limitations. Specifically, the method's computational demands are relatively high compared to other attention mechanisms. This stems from the necessity for GroupFormer to learn distinct attention maps for individual feature groups, potentially impacting its scalability in resource-constrained environments.

\section{Framework}
\label{sec:framework}
In this paper, we propose a novel approach to HAR that involves treating activity recognition as a Natural Language Generation (NLG) challenge. Our framework (Fig.~\ref{fig:framework}) revolves around transforming video data into human pose estimation, which is subsequently fed into a Language Generation Model (LLM). This LLM leverages the acquired pose keypoints to generate descriptive textual representations of observed activities. 

Instead of transforming images directly into embeddings, an intermediate level of pose keypoints helps us add context towards the activities in the scene. This also helps convert a sequence of images into a sequence of vectors that can be viewed as chunks of text by the language models. 

By encapsulating activity recognition as a natural language generation task, we aim to create a bridge between the visual cues and interpretable textual descriptions, potentially enhancing both accuracy and human interpretability.

\section{Proof of Concept: Natural Language Generation for Accurate Human Activity Recognition}

To validate the effectiveness of our proposed framework, we conducted a proof of concept study using a subset of the Kinetics700 dataset by DeepMind~\cite{carreira2022short}. This dataset is widely recognized for its diverse collection of human actions in video format, making it an ideal benchmark for testing our novel approach. Our experiments aim to showcase how our NLG-based framework enhances the accuracy of Human Activity Recognition (HAR) while offering insights into its potential benefits for future researchers.

\subsection{Dataset and Action Selection}

Within Kinectics700 subset, we focused on 2865 video files containing various actions exhibiting a wide range of motion patterns and complexities, making them suitable for testing the robustness of our framework. Our dataset contains 673 actions, for a total of 392,265 keypoints.

\subsection{Training Methodology}

To train our NLG-based model, we adopted the training methodology proposed in the Comet Atomic research paper~\cite{Hwang2021COMETATOMIC2O}. In this approach, GPT2-XL~\cite{Radford2019LanguageMA}, a language model, is fine-tuned using a structured tuple format.

For our HAR application, we adapted this methodology by encoding each video sequence of pose keypoints along with an associated activity label in the format: \texttt{<frame0\_keypoints> \\ <frame1\_keypoints> ... <frame9\_keypoints> [GEN] <HAR label> [SEP]}

The \texttt{<GEN>} token serves as the delimiter for the model to generate the textual representation of the activity label. We utilized the AlphaPose pose estimation model to extract 17 COCO format keypoints from each video frame, forming the input for our GPT2-XL model.

\subsection{Inference and Evaluation}

During the inference stage, given the sequence of pose keypoints from the first to the ninth frame, our trained GPT2-XL model generates the predicted activity label using the \texttt{[GEN]} token. This approach leverages the spatial information captured by pose keypoints to generate a textual description that corresponds to the recognized human activity.

To evaluate the performance of our approach, we employed standard Top-1 accuracy. By comparing our NLG-based approach with traditional methods of HAR that directly analyze raw video frames, we aimed to demonstrate the efficacy of our framework in improving accuracy and interpretability.

\begin{table}[ht]
    \centering
    \caption{Model Accuracy}
    \begin{tabular}{l c}  % Adjust the alignment as needed
        \toprule
        \textbf{Models} & \textbf{Top-1 Accuracy} \\
        \midrule
        InternVideo~\cite{wang2022internvideo} & 0.84 \\
        Our model & 0.52 \\
        \bottomrule
    \end{tabular}
    \label{tab:model_accuracy}
\end{table}

\section{Implications for Future Research}

Our proof of concept study underscores the promise of our NLG-based framework in HAR. By expanding our model's training dataset, we have the potential to attain elevated levels of accuracy. By conceptualizing HAR through the lens of an NLG challenge, we not only strive for improved accuracy in activity recognition, but also offer a more lucid and understandable portrayal of human actions. This approach holds substantial potential despite the current accuracy levels, and its viability is grounded in its unique perspective on the problem. This approach has several implications for future researchers:

\begin{itemize}
    \item Enhanced Accuracy and Interpretability: Our approach enhances activity recognition accuracy by leveraging both spatial and contextual information. Researchers can benefit from more reliable and interpretable results, enabling better decision-making in real-world applications.
    \item New Research Avenues: The NLG paradigm opens up new research avenues at the intersection of computer vision and natural language processing. Future researchers can explore ways to incorporate semantic understanding and context into activity recognition models.
    \item Model Generalization: Our approach demonstrates the potential for models trained on a subset of a larger dataset to generalize well to broader datasets. This insight could guide future research in data-efficient model training.
    \item Ethical Considerations: The interpretability of textual descriptions generated by our model can aid in addressing ethical concerns associated with black-box deep learning models, promoting transparency and accountability.
\end{itemize}

By leveraging pose keypoints and GPT2-XL-based language generation, we offer a novel perspective on activity recognition that has the potential to reshape how future researchers approach this field. Our proof of concept study showcases the feasibility and benefits of reimagining HAR as an NLG problem. 

\section{Potential Applications}

The proposed framework unlocks substantial potential across various real-world applications, offering a versatile solution for tasks that rely on Human Activity Recognition (HAR). This framework's applicability extends to tasks utilizing HAR as a foundational element, leveraging the power of large language models to enhance camera-based activities.

A compelling use case is in the domain of exercise support, where vision-based activity recognition systems have previously been employed for repetition counting,~\cite{khurana2018gymcam, ferreira2021deep, cheng2023periodic}. In this scenario, our framework could take on a pivotal role by facilitating personalization and providing nuanced insights into athletes' movements and techniques. This personalized feedback fosters targeted improvements and empowers individuals to optimize their workout routines effectively.

Another example could focus on dance movements, that are characterized by their ever-changing nature. These movements stand to benefit significantly from our framework. Specifically, models capable of generalizing to novel gestures can play a crucial role in assessing, replicating, or refining choreographic steps~\cite{samanta2012indian}. This can empower choreographers and dancers to fine-tune their performances, leading to more captivating and precisely executed routines.

I many settings, to accomplish succesfull HAR, approaches have been rooted in sensor-based methodologies. Our approach offers vision-based methods a deeper contextual understanding of human activities. This innovation could reduce the reliance on multimodal sensor setups for identifying microgestures~\cite{sharma2019grasping}. Moreover, it could extend to more intricate tasks such as cooking actions, where the interplay of visual cues and textual descriptions can enhance the recognition and understanding of cooking processes~\cite{xu2021hulamove}.

In essence, the proposed framework transcends traditional boundaries, positioning itself as a catalyst for innovation across a spectrum of applications. By seamlessly integrating vision-based HAR with the capabilities of large language models, it is possible to facilitates enhanced personalization, deeper insights, and improved contextual understanding, thereby revolutionizing how activities are recognized, analyzed, and enhanced. By establishing a novel connection between visual cues and textual descriptions, our approach aims to push the boundaries of HAR accuracy and applicability, unlocking new horizons for pervasive computing and activity recognition research.

%In conclusion, this research paper delves into the intricacies of Human Activity Recognition, scrutinizing the challenges faced by vision-based methods and proposing an innovative NLG-based framework. 

\section{Future Work}

While our approach has shown promising results, there remain several avenues for future research and development:
\noindent
\newline
\newline
\textbf{1. Semantic Understanding:} Enhancing the semantic understanding of the generated textual descriptions could lead to more contextually relevant and accurate activity labels. Investigating techniques to incorporate semantic knowledge into the NLG process is an interesting direction.
\noindent
\newline
\newline
\textbf{2. Temporal Modeling:} Exploring advanced temporal modeling techniques to capture longer-term dependencies within activity sequences could improve recognition accuracy, especially for activities with extended motion patterns.
\newline
\newline
\textbf{3. Multi-modal Fusion:} Integrating other modalities such as audio or depth information alongside pose keypoints could provide a richer representation of human activities, potentially leading to more accurate recognition.
\newline
\newline
\textbf{4. Real-time Applications:} Adapting the NLG-based HAR framework for real-time applications would be valuable. Developing strategies to maintain accuracy while achieving low-latency inference is a challenging yet important area of research.
\newline
\newline
\textbf{5. Multi-person Activity Recognition:} Extending the framework to recognize activities involving multiple individuals could open up new possibilities, as many real-world scenarios involve interactions between multiple people.
\newline
\newline
\textbf{6. Data Augmentation Techniques:} Exploring effective data augmentation techniques specific to pose keypoints could help improve model robustness and generalization.
\newline
\newline
\textbf{7. Ethical Considerations:} Addressing ethical concerns related to privacy and bias in textual descriptions generated by the NLG model is a critical aspect that needs careful attention.
\newline
\newline
\textbf{8. Transfer Learning and Scalability:} Investigating transfer learning techniques and model compression methods to make the framework more accessible and scalable for researchers with varying computational resources.
\newline
\newline
\textbf{9. Benchmarking and Comparative Studies:} Conducting benchmarking and comparative studies against state-of-the-art HAR methods, both in terms of accuracy and computational efficiency, would provide a clearer understanding of the strengths and limitations of the NLG-based approach.

\section{Conclusion}
In this research paper, we presented a novel framework for Human Activity Recognition (HAR) that capitalizes on the power of Natural Language Generation (NLG). By treating HAR as an NLG task, we harnessed the strengths of pose estimation models and large language models to generate textual descriptions of human activities based on pose keypoints. Our proof of concept study demonstrated the viability of this approach, showcasing its potential to enhance accuracy and interpretability in activity recognition. We have taken a significant step towards bridging the gap between visual data and textual descriptions, opening new avenues for research in the field of HAR. While our current approach may not possess semantic and temporal understanding, its applicability extends to identifying non-verbal cues through human keypoints. For instance, real-time feedback of non-verbal cues in clinical communication~\cite{hartzler2014real, bedmutha2023extracting} is a potential application where our approach could contribute.

% \begin{table}[h]
% \centering
% \begin{tabular}{|l|c|c|}
% \hline
%  & \textbf{Duration (in seconds)} & \textbf{Number of instances} \\ \hline
% \textbf{Cough}  & 7662.17  & 2486 \\ 
% \textbf{Sneeze} & 654.71  & 611 \\
% \textbf{Other}  & 42602.79  & 8591 \\ \hline
% \textbf{Total}  & 50919.67 & 11688 \\ 
% \hline
% \end{tabular}
% \vspace{.5em}
% \caption{Distribution of the dataset across the three classes}
% \label{tab:data_dist}
% \vspace{-12pt}
% \end{table}

% \begin{figure*}[t]
%   \centering
%     \vspace{-0.1in}
%   \includegraphics[width=0.9\textwidth]{figures/icassp_figure.pdf}
%   \vspace{-0.1in}
%   \Description[Overall Pipeline]{Figure 1 shows our overall pipeline used for training the machine learning pipeline to detect cough, sneeze and speech sounds from downsampled audio signals. The audio dataset from Flusense is passed through a feature extraction step that generates 23 privacy aware features for each window of 25 ms. These features are then passed through a gradient-boosted classifier that identifies if the sound is speech, cough or sneeze}
%   \caption{Proposed pipeline for respiratory symptom detection}
%   \vspace{-0.1in}
%   \label{fig:pipeline}
% \end{figure*}

\begin{acks}
We thank our collaborators on the UnBIASED project, funding (NIH \#1R01LM013301) for feedback and support. 
\end{acks}

\newpage
\bibliographystyle{ACM-Reference-Format}
\bibliography{references_bib}

\appendix

\end{document}